\documentclass[a4paper,12pt]{article}
\usepackage[utf8]{inputenc}
\usepackage{latexsym}
\usepackage{amssymb}
\usepackage{verbatim}
\usepackage{graphicx}
\usepackage{epstopdf}

\usepackage[T1]{fontenc}
\usepackage[utf8]{inputenc}

\usepackage[british]{babel} 
\usepackage{csquotes}        
\usepackage[
    notes,
    backend=biber,
    hyperref=true  
]{biblatex-chicago}

\usepackage{xcolor}

\title{Towards a Measure Theory of Semantic Information}
\author{George M. Coghill \\  University of Aberdeen \& \\ Edinburgh Theological Seminary. \\ email: g.coghill@abdn.ac.uk \&  gcoghill@ets.ac.uk}
 \date{}

\begin{document}
\maketitle

\begin{abstract}

A classic account of the quantification of semantic information is that of Bar-Hillel and Carnap. Their account proposes an inverse relation between the informativeness of a statement and its probability. However, their approach assigns the maximum informativeness to a contradiction: which Floridi refers to as the Bar Hillel-Carnap paradox. He developed a novel theory founded on a a distance metric and  parabolic relation, designed to remove this paradox. Unfortunately his approach does not succeed in that aim. 

In this paper I critique Floridi’s theory of strongly semantic information on its own terms and show where it succeeds and fails. I then present a new approach based on the unit circle (a relation that has been the basis of theories from basic trigonometry to quantum theory). This is used, by analogy with von Neumann's quantum probability, to construct a measure space for informativeness that meets all the requirements stipulated by Floridi and removes the paradox. In addition, while contradictions and tautologies have zero informativeness, it is found that messages which are contradictory to each other are equally informative. The utility of this is explained by means of an example.

\end{abstract}


\section{Introduction}

There is a growing interest in Information and its philosophy, particularly with the current wave of agitation (positive and negative) surrounding the successes of AI across a range of fields. Associated with this there has been interest in measuring information, and informativeness. These include Shannon information, \autocite{shannon49} and the work of MacKay, \autocite{mackay69} but they were not focussed on semantics. In fact, Shannon explicitly stated that his quantification was purely syntactic. On the other hand, meaning is also deemed to be important, or even fundamental, in the study of information, therefore there have been methods devised to quantify the semantic content of information, the best known of which is by Bar-Hillel and Carnap. \autocite{barcar53}. Not surprisingly, more recently there have been efforts to merge these (and other) understandings and uses of the term ``information'' into a single framework. \autocite{burgin10}  

A different view is presented by Floridi, who has developed a comprehensive philosophy of information based primarily on semantic information. He restricts the term ``Information'' to meaningful statements and suggests referring to other types by the term ``Mathematical Theory of Communication'' (MTC). \autocite{floridi11} Within that context he has proposed a novel quantification method  to address a putative issue with Bar-Hillel and Carnap's  formulation, namely that their approach has a contradiction being maximally informative. This he referred to as the Bar-Hillel--Carnap paradox (BCP).  He hoped his method would provide a solution to the BCP and provide a robust foundation for the quantification of semantic information. He named his theory  the ``Theory of Strongly Semantic Information'' (TSSI).\autocite{floridi04a} There have been a number of engagements with, and critiques of, TSSI.\footnote{For example: Cevolani \autocite{cevolani14}, D'Alfonso \autocite{dalfonso11}, Ferguson \autocite{ferguson15}, and Wolff \autocite{wolf11}} However each of these has assumed that Floridi’s construction and analysis is correct in what it asserts and so no one assessed whether what he has developed actually achieved its aim. Their actual criticisms  mainly focussed on what followed therefrom. 

The focus of this paper is a critical analysis of TSSI. To that end I review TSSI on its own terms, as it is presented by Floridi. We shall find that there are a number of issues that reduce its usefulness with regard to the task for which it was designed. Nonetheless, it provides a step in the right direction, and does suggest a way forward. Having identified the problems with TSSI, in the rest of the paper I propose an improved version of TSSI that overcomes the identified deficiencies and provides a more general means to measure the informativeness of a message, which I shall refer to as the ``Measure Theory of Semantic Information'', (MTSI). In particular it meets all Floridi's criteria for the quantification of information and removes the BCP.

\section{Information and the ``Bar-Hillel--Carnap'' paradox. }

Floridi, following Dretske, \autocite{dretske99} adheres to the Veridicality Thesis (VT). That is, in order to count as information a message must be true.\autocite[chap. 4]{floridi11} The idea of ``false information'' is in the same boat as the term ``false friend'': a useful and evocative description, but with the clear meaning of not being a friend at all.\autocite{dretske99} This view is not universally accepted, but because the focus of this paper is TSSI (and VT is essential to that) it will, for our purposes and without prejudice, be taken as read that VT holds. We can then focus on Floridi's arguments for the quantification of information in TSSI.

Floridi borrows from situation logic the term {\em infon} (symbolised as $\sigma$) \autocite{devlin91, israel90} ``to refer to discrete items of factual semantic information qualifiable in principle as true or false, irrespective of their semiotic code and physical implementation''\footnote{For the purposes of this paper these may be considered to be {\em propositions} as has been done by \autocite{cevolani11} and \autocite{dalfonso11}. However, while keeping that in mind I shall continue to use the term ``infon'' throughout this paper for continuity with Floridi.}. 
Using this, he presents the General Definition of Information (GDI) as a quad-partite definition:

\vspace{8mm}
\noindent $GDI_\sigma$ (an infon) is an instance of semantic information iff: \\
\begin{itemize}
\item GDI.1:	$\sigma$ consists of $n$ {\em data} (d) for $n \geq 1$ \\
\item GDI.2:	the data are {\em well-formed} ($wfd$) \\
\item GDI.3:	the wfd are meaningful ($mwfd = \delta$) \\
\item GDI.4:	the $\delta$ are {\em truthful}.
\end{itemize}

This definition makes use of a prior term {\em data}, or {\em datum} which is ultimately a {\em lack of uniformity}  \autocite[p 23]{floridi10}. Which leads to a description of information as ``a distinction that makes a difference.''\footnote{This description is attributed, by Floridi and others, to MacKay \autocite{mackay69}. However neither I nor anyone else, including MacKay's widow has been able to confirm that he is the originator of the description, nor where it might be found. However, insofar as it serves as a the contradictory to the common description of being uninformative as ``a distinction without a difference'' it remains an accurate description and should continue to be used.} The definition of a datum is then:\autocite[p 23]{floridi10}

\begin{itemize}

\item[Dd:] {\em datum} $=_{def}$ $x$ being distinct from $y$, where $x$ and $y$ are two uninterpreted variables and relation of `being distinct' as well as the domain, are left open to further interpretation.

\end{itemize}

Floridi uses these to distinguish between the Bar Hillel and Carnap approach, that he calls the ``Theory of Weakly Semantic Information'' (TWSI), in which truth merely supervenes on information, and gives rise to a paradox, and the ``Theory of Strongly Semantic Information'' (TSSI), his own formulation which he believes overcomes that paradox. The former is defined by GDI.1 -- GDI.3, whereas the latter uses all four in the definition of information.

Floridi provides an analysis of where TWSI leads, \autocite{floridi11} in particular as described by Bar-Hillel and Carnap \autocite{barcar53} as a grounding for his TSSI. The context for Floridi's approach is as a response to the definition of semantic information put forward by Bar-Hill and Carnap in the 1960s. They identified the amount of information in, or the semantic content (CONT) of, a proposition as being inversely related to the prior probability of the proposition (referred to as the Inverse Relationship Principle, IRP). 

$$CONT(p) = 1 - P(p)$$
 
\noindent Here $P(p)$ is the prior probability of $p$ and $CONT$ could be, for example a set of possible worlds, or a set of propositions, and is a measure of the likelihood of p not being the case. One upshot of this is that while the semantic content of a tautology is zero, as one would expect, $CONT(\perp)$, is 1. That is, a contradiction is maximally informative. This is rather surprising. As Bar-Hillel and Carnap put it:

\begin{quote}
``It might perhaps, at first, seem strange that a self-contradictory sentence, hence one which no ideal receiver would accept, is regarded as carrying with it the most inclusive information. It should, however, be emphasized that semantic information is here not meant as implying truth. A false sentence which happens to say much is thereby highly informative in our sense. Whether the information it carries is true or false, scientifically valuable or not, and so forth, does not concern us. A self-contradictory sentence asserts too much; it is too informative to be true.'' \autocite[p 229]{barcar53}.
\end{quote}

That is, Bar-Hillel and Carnap reject the VT and see truth as supervening on information.  It is this feature that gives rise to what Floridi calls the {\em Bar-Hillel--Carnap paradox} (BCP), wherein the impossible situation of a contradiction is the most conveys that maximum information.

\section{The Theory of Strongly Semantic Information}

In order to understand the criticism that will follow in the next section we need to present an outline of the substance of Floridi's TSSI. 
Floridi proposes that the best place to start is with ``an  analysis of the quantity of semantic information in $\sigma$ including a reference to its alethic value. This is TSSI'' \autocite[p 117]{floridi11}. 
To this end he identifies three desiderata for a theory of semantic information; it should: 
\begin{enumerate}
    \item[D.1:] avoid any counterintuitive inequality comparable to BCP;
    \item[D.2:] treat the alethic  value of $\sigma$ not as a supervenient but as a necessary feature of semantic information, relevant to the quantitative analysis;
    \item[D.3:] extend a quantitative analysis to the whole family of information-related concepts: semantic information vacuity and inaccuracy, informativeness, misinformation (what is ordinarily called `false information'), disinformation.
\end{enumerate}
These three desiderata are fundamental. The first two are unproblematic; the third is unproblematic in the abstract, but the way it is unpacked may give rise to some issues (see Section \ref{probs}). It this unpacking that Floridi proceeds to develop.  Throughout the discussion Floridi makes use of an example universe, $E$, consisting of all possible worlds, or states,  arising from the conjunction of a set of basic infons.\footnote{Floridi presents it as containing all possible {\em messages}, but as D'Alfonso \autocite{dalfonso11} has pointed out, this is not the case.} 

As well as the statement of BCP as an inverse relation: 

$$CONT(\sigma) = 1 - P(\sigma)$$

\noindent based on what an infon excludes, he also highlights the relation between the informativeness, $\iota$, of an infon and its semantic content, stated as: 

$$\iota(\sigma) = CONT(\sigma)$$

The particular quantitative measure that Floridi chooses for TSSI is not probability but the discrepancy, $\vartheta$, from the actual state of affairs.\footnote{Floridi gives a fuller set of comparisons to justify the development of TSSI, but this summary is sufficient for the purposes of this paper.} 

As a step towards formulating a relation to enable the calculation of informativeness for an infon, Floridi identified five criteria that he considered any method of quantification must meet (which he labelled M.1-M.5). These criteria are uncontroversial given the aims of TSSI, but, as we shall see in the next section Floridi's chosen means of implementation does not actually meet all these criteria.

The criteria are as follows: the true state will have zero discrepancy [M.1]; both a tautology [M.2] and a contradiction [M.3] will have the maximum discrepancy; contingently true [M.4] and contingently false [M.5] infons will have discrepancies that lie strictly in the range between zero and the maximum discrepancy.
\footnote{For details of the formal expression of these five criteria see \autocite[p 120f]{floridi11}.}

\subsection{Degrees of Inaccuracy}

The model universe is maximal for states\footnote{Though, as noted, not for messages.} (i.e. no further conjunctive propositions may be added without introducing a contradiction) and so it must contain the true state.  Since the infons are conjunctions, any infon, $\sigma$, other than that representing the actual state of affairs will be false. However, different infons will contain greater or fewer components ({\em conjuncts}) that are false, and this gives rise the idea of {\em degree of falsity} or {\em inaccuracy}. It is straightforward then to create a metric of the distance from the true state of affairs as the ratio of the number of false components to the length of the infon (that is, the total number of conjuncts in the infon). More formally: 

\begin{equation}
-\vartheta(\sigma) = -e(\sigma)/l(\sigma)
\label{inac}
\end{equation}

Here $l$ is the length of the infon (which in the case of Floridii's example is 6), $e$ is the number of erroneous conjuncts in the infon, and $\vartheta$ is the distance of the infon from the true state of affairs  (the negative sign refers to the fact that it deals with degrees of error); and spans the range 0 (matching the actual situation) to -1 (the infon contains no truth). 

\subsection{Degrees of Vacuity}

The other aspect of distance from the actual state of affairs arises when the infon is true but more abstract than the actual (true) state. The most extreme example of this is a tautology, which includes the actual situation, but is uninformative due to its being true in all circumstances. This has a distance of +1. In order to fill in the gap between these two extremes Floridi introduced the semi-dual.\footnote{A semi-dual is an infon in which the operators are changed from conjunction to disjunction, but the components are not negated. For example a contradiction is the semi-dual of a tautology.}  By this means a set of classes are identified which contain all the infons with the same number of disjunctions. Each member of the class will have the same number of ways of being true given that some components of the infon are false. So in this case the distance is the ratio of the number of ways of being true, $n$, to the size of the universe\footnote{Here $s$ represents the number of values an infon can take (in this case two: \{T, F\}), and $l$ is the length of the infon} $s^l$ (where $l$ is the length of the infon and $s$ is the number of values the infon can take: 2 in this case). More formally:  

\begin{equation}
\vartheta(\sigma) = n/s^l
\label{dist}
\end{equation}

\subsection{Degrees of Informativeness}

Now that a putatively suitable distance metric has been identified for these two situations Floridi is able to use it to provide a means of quantifying the degree of informativeness, $\iota$, of an infon. He proposes that the distance be viewed as spanning the range [-1, +1], with the actual state of affairs at the origin, the LHS being the degree of inaccuracy (hence the negative value) and the RHS the degree of vacuity. 

The fact that, as Floridi sees it,  the BCP arises, at least in part, from the inverse relation between information content and probability in that formulation suggests that a relation that identifies the actual state of affairs as having maximum informativeness and the two extremes (tautology and contradiction), zero informativeness will serve better as a metric.

\begin{figure}[btph]
\centering
\includegraphics[scale=1.1]{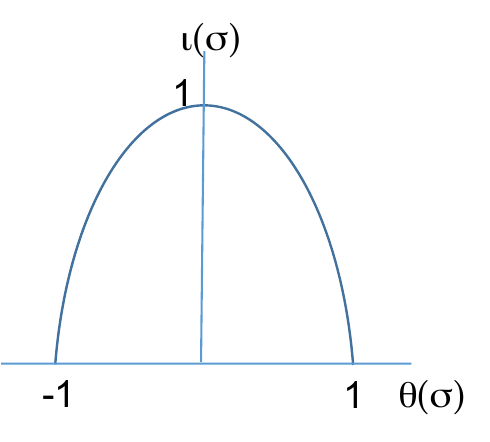}
\caption{\label{if} The degree of informativeness (after Floridi \autocite{floridi11}). Here, $\iota(\sigma) = 1 - \vartheta^2(\sigma)$}
\end{figure}

He proposes a quadratic relation as being the simplest relation that meets a number of criteria that he considers mandatory for a function to quantify the degree of informativeness. Most of these are unobjectionable, but a couple appear more problematic and less than optimal. (We will describe thee and the issues surrounding them in section \ref{contra} below). The precise formulation used by Floridi is: 
\begin{equation}
\iota(\sigma) = 1 - \vartheta^2(\sigma)
\label{inf}
\end{equation}

\noindent see Fig. \ref{if}.

The motivation, in part at least, for this quadratic is the claim that it meets a number of other conditions that he believes follow directly from M.1- M.5. These conditions (which he labels E.1 - E.6) are: that when the discrepancy $\vartheta$, is zero the informativeness of the infon, $\iota$, is 1 [E.1]; that the integral of $\iota$. must be bounded in the interval between minimum and, maximum values (i.e. it must be a proper integral) [E.2]; when the discrepancy is at a maximum, the informativeness must be zero [E.3]; when the discrepancy is in the range between minimum and maximum, then the informativeness lies in the range $0 \ldots 1$ [E.4]; a small variation in discrepancy yields a substantial change in informativeness; [E.5]; and finally [E.6]: the marginal information function (which is the derivation of the equation for informativeness) is a linear function. This last is meant to approximate the ideal that ``all atomic messages ought to be assigned the same potential degree of informativeness''.
It can be agreed that these are, for the most part, suitable criteria and conditions for any function that aims to quantify informativeness to possess. Unfortunately, as we shall see in the next section TSSI fails to meet several of them. Most importantly it does not yield an informativeness of zero for a contradiction, and therefore fails in the central aim of removing the BCP.

\section{\label{probs} Some problems with TSSI}

It is clear that Floridi has taken a  step forward in helping to understand semantic information and the various desiderata associated therewith. However,  there are a few problems in the current formulation that limit its applicability. 

\subsection{\label{contra} The status of Contingency and Contradiction}

A major reason, if not {\em the} reason, for the development of TSSI was to provide a solution to the BCP. That being the case, having a clear and coherent means of calculating the informativeness of a contradiction is crucial, and it is this that the relation in Equation \ref{inf} seeks to provide.

Floridi \autocite[p 120]{floridi11} gave a formal statement of each of the criteria; I state the problematic ones here:

\begin{equation}
(M.3) ~~~\mid\models_{\neg \exists w} \sigma \rightarrow f(\sigma) = -1
\end{equation}

That is, ``if (it is estimated that) $\sigma$ is false and conforms to no possible situation, then $\sigma$ is a contradiction and it is assigned the maximum degree of negative discrepancy.''\autocite[p 120]{floridi11} This is as it should be and so the calculation of inaccuracy should naturally give this answer for a contradiction, but, as we shall see, that is not what happens. 

The next criterion is stated as: 

\begin{equation}
(M.4) ~~~\mid\models_{w -\vartheta} \sigma \rightarrow (0 > f(\sigma) > -1)
\end{equation}

\noindent meaning ``if (it is estimated that) $\sigma$ is contingently false, then it is assigned a degree of discrepancy with a value less than 0 but greater than --1 (degrees of semantic inaccuracy).''\autocite[p 120]{floridi11} 

Unfortunately for any state that is a conjunction of atomic infons, the distance metric given in equation \ref{inac}, is in violation of both these criteria. For example if the worlds/ situations are conjunctions of $n$atomic infons giving a universe, E, of $2^n$ possible worlds/situations.\footnote{Floridi uses states consisting of six atomic infons, giving a universe of 64 worlds/situations.}  The distance from the actual world, $w$, is calculated from equation \ref{inac}. 

In the case where all the atomic infons in the state are false, the degree of inaccuracy (i.e. the distance from $w$), $\vartheta$ is $\frac{-n}{n}$, which is equal to -1. Since this is a contingent falsehood it is in violation of (M.4). 

On the other hand, if we apply Equation \ref{inac} to a contradiction then we will find that the inaccuracy is not equal to -1. This is because any contradiction must contain at least one conjunct that is true, and so its degree of inaccuracy can never be -1 (in violation of (M.3))! In fact for a contradictory infon containing $n$ conjuncts the number of true conjuncts will lie in the range from 1 to $n-1$, and so $\vartheta$ will lie in the range $\frac{1}{n}$ to  $\frac{n-1}{n}$.\footnote{This also gives rise to an issue highlighted by D'Alfonso \autocite{dalfonso11} relating to inconsistency. He identifies that inconsistent infons that have greater or fewer true constituents would have higher or lower informativeness. Unlike Floridi, and the position proposed in this paper, he thinks that is intuitively as it should be.}

In the light of both these issues we will need to find a more appropriate means to represent the inaccuracy and vacuity of an infon. I address this in section \ref{gen}.

\subsection{\label{infoS} The Information Space}

Another issue with this representation arises if we wish to use the informativeness relation to represent the relationship between different information statements, or shifts between infons (e.g. by means of abstraction). As an illustration consider the situation where the infon (length six as in the original example) contains one false atomic statement. In that case the inaccuracy would be -0.167 (from equation \ref{inac}). If we now abstract the infon, by means of a single disjunction, it becomes true with a vacuity of +0.047 (from equation \ref{dist}). This is discontinuous in the most basic sense: the single operation makes the deviation, and hence the informativeness jump from left hand side of the information space to the right hand side without passing through zero (and to jump between different metrics). And this jumping back and forth would continue (and potentially get larger) as more falsehoods and disjunctions are introduced into the infon. This suggests that it is worth exploring to see if there is a representation and relationship that better captures what Floridi is aiming for here.

Another problem is that it is not clear what purpose [E5] serves. It is not unique to a quadratic (it can be equally well met by a circle), and it arguably conflicts with [E6]. Therefore it seems best to drop it as a criterion, and I shall ignore it for the rest of this paper.

There have been a number of critiques and criticisms of Floridi's approach \autocite{cevolani14, dagostino13, dalfonso11, crukovic08, wolf11}. Some have made interesting points, but often do not hit the mark with respect to the space. Several of these have been dealt with by Sequoiah-Grayson \autocite{sequoiah-grayson07}. 

\subsection{Some conclusions regarding TSSI}

In this section we have highlighted a number of issues, and criticisms of Floridi's approach, in particular, it does not achieve the main aim of resolving the BCP. At this point it is worthwhile to stop and reflect on the positive contributions and shortcomings of TSSI. There can be summarised as follows:

On the positive side: 

\begin{itemize}

\item Floridi identifies the weakness of TWSI, and the undesirable nature of the relation that makes a contradiction the most informative of messages.

\item He also provides a set of desiderata that provide, in a very clear way, the ideals for a theory of semantic information.

\item The quantitative relation that he suggest is a step forward in that it points towards a means of calculating the informativeness of an item without having to assign non-zero informativeness to a contradiction (while not quite achieving that goal).

\end{itemize}

However there are a number of issues which highlight that, despite these contributions, TSSI is not fit for purpose as it stands

\begin{itemize}

\item The RHS and LHS of the space do not have the same metric, which is not desirable for a single straight line axis.

\item A contingent message which has only false atoms will have an informativeness of zero, in violation of M3. 

\item A contradiction, because by definition it will have at least one true atom must have a non-zero informativeness. This is the main problem because it violates both M3, and the main motivation for the work, and therefore does not fully solve the problem of BCP.

\end{itemize}

Given the fundamental nature of these issues there is scope to explore another representation for a space in order to comply with the positive contribution while addressing the problems.

\section{\label{gen} A more general representation of Semantic Information}

In this section I shall propose a different representation for the Information Space (IS) that will provide the means to meet the desiderata identified by Floridi and, most importantly, remove the BCP. This can be done utilising methods that have been used very successfully in other domains. In the Mathematical Theory of Communication, and related areas (such as Control Theory and Signal Theory) the Unit Circle (UC) is often used as a convenient means of representing signals and analysing stability. This is not surprising since in these domains sinusoids and trigonometric relations are commonplace, and they are defined in relation to a circle. In addition, in developing Quantum Probability (QP), based on Hilbert spaces, von Neumann utilised similar methods.\autocite{neumann32} This being the case I argue that, given the problems identified in the previous section, the UC is worth exploring in this context as a better means of representing the relations of quantification of information. 

It is a commonplace of such domains that incompatible variables or ranges be represented by orthogonal axes, whether it be Real and Imaginary numbers in communication theory or `up' and `down' in quantum theory. Two key features of the UC  that are relevant here are the ease with which one can calculate quantitative relations by means of the cosine rule, and related to that, the fact that pure True infons and pure False infons are orthogonal to each other (and hence with no mapping since the cosine is zero).  As such, True and False can be used as the axes of the space with respect to which the deviations from the actual state of affairs are measured (we can label these axes $\vartheta_T(\sigma)$ and $\vartheta_F(\sigma)$). Here the true state-of-affairs is again the origin.
Messages which contain both true and false atoms will be distributed in the space between the axes.

\begin{table}
\caption{\label{tab} The complete set of messages. The representation used here is from digital systems, where conjunction is shown as a product and disjunction as a sum, this is equivalent to, but slightly more compact than, the usual normal forms in symbolic logic.}
\begin{tabular}{|c|c|c|c|c|}
\hline
Message & Message & No. of worlds in & No. of true atoms \\
Number && which it is True &  \\
\hline
M0 &  $xx'$ & 0 &1  \\
\hline
M1 & $xy$ & 1 & 2  \\
\hline
M2 & $xy'$ & 1 & 1  \\
\hline
M3 & $x'y$ & 1 & 1  \\
\hline
M4 & $x'y'$ & 1 & 0  \\
\hline
M5 & $x$ & 2 & 1   \\
\hline
M6 & $y$ & 2 & 1  \\
\hline
M7 & $xy + x'y'$ & 2 & 2  \\
\hline
M8 & $xy' + x'y$ & 2 & 2  \\
\hline
M9 & $y'$ & 2 & 0  \\
\hline
M10& $x'$ &2 & 0  \\
\hline
M11 & $x + y$ & 3 & 2  \\
\hline
M12 & $x + y'$ & 3 & 1  \\
\hline
M13 & $x' + y$ & 3 & 1  \\
\hline
M14 & $x' + y'$ & 3 & 0  \\
\hline
M15 & $x + x'$ & 4 & 1  \\
\hline
\end{tabular}
\end{table}

\begin{figure}[btph]
\centering
\includegraphics[scale=0.8]{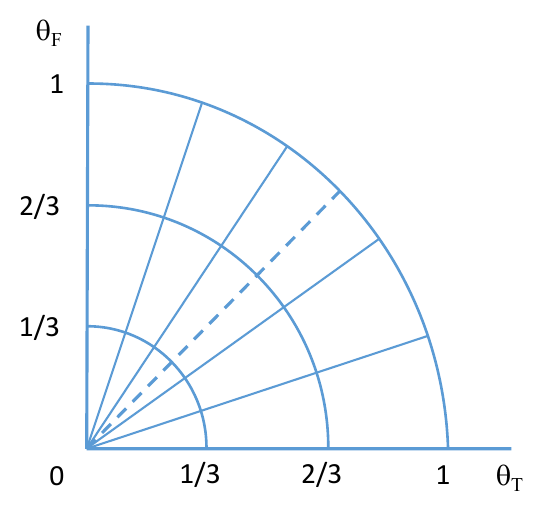}
\caption{\label{im} The structure of the information metric space.}
\end{figure}

The first things that need to be done then are to change the space from that shown in Fig \ref{if} (2-D) to that shown in Fig \ref{im} (3-D, with the third dimension, representing $\iota(\sigma)$, coming out of the page). Then we need to decide on what the best, and most coherent, means to calculate the distance from the true state-of-affairs is. In deciding what scales to utilise in measuring the distance from the true state, we need to consider two things: the distance along each axis, and the position in the plane. Before we settle this we should look at what messages can be created. This is best done by means of an example. Floridi used a universe consisting of three atoms and two predicates. which gave a space of 64 states (or worlds). However, as D'Alfonso \autocite{dalfonso11} has pointed out, the actual number of possible messages is $2^{64}$ ($10^{19}$) for this universe. We wish our space to represent all these messages and hence will need a simpler example to demonstrate things. In fact the only manageable possibility for these purposes is a single predicate with two atoms. This yields a universe of two atomic infons, four worlds (or states), and sixteen messages.\footnote{Even three atomic infons would give eight worlds and 256 messages; which is too unwieldy for illustrative purposes. To put this in perspective, consider a high quality print of this page with, say, 600 dpi. That would give around 28 million dots on the page: several orders of magnitude fewer than would be required to represent all the possible messages in Floridi's original example. In fact, about a decade ago it was estimated that around $10^{19}$ pieces of information had been generated since the invention of writing.}
Table \ref{tab} shows the complete set of messages for the two atomic infons $x$ and $y$.\footnote{A note about nomenclature. In this paper, I am using the symbolic representation from digital systems and communication for ease of presentation That is, conjunction is represented by a dot (though it is usually omitted where no confusion will ensue). Disjunction is represented by a plus sign and negation by a prime stroke. This gives rise to the Sum-of-Products (SOP), equivalent to The Conjunctive Normal Form (CNF) of propositional logic.} From Table \ref{tab} we see that the number of worlds in which any message may be true (or false) ranges between 0 and 4. In particular note that the tautology is true in 4 worlds (and the contradiction is false in 4 worlds). 

Again we wish to quantify the deviation $\vartheta$ of each infon from the actual state.
Let us assume that the actual state of affairs is, for example, $xy$. That state is true in one world, so in order to calculate the deviation we first identify the difference in the number of worlds between the state of interest and the true state, call that $k$, (e.g. the difference between $xy$ and $T$ is 3 for this example). 
The deviation from the true state (for the horizontal axis) is then the ratio of $k$ to that maximum difference, which is the  total number of worlds (call that $m$) minus 1. This is expressed formally as:

\begin{equation}
\vartheta_T(\sigma) = \frac{k}{(m - 1)}
\label{mtsiT}
\end{equation}

This is similar to the metric used by Floridi for the RHS of his space. We now require two variables to represent the deviation from the true state of affairs; call these $\vartheta_T(\sigma)$ and $\vartheta_F(\sigma)$ for the {\em True} and {\em False} axes respectively. To illustrate this consider the $\vartheta_T(\sigma)$ axis. Here $xy$ is the putative actual state of affairs. Then, if that is the case, both $x$ and $y$ must be true, but more vacuous (from \ref{mtsiT}), similarly $x+y$ must also be true but even more vacuous. All these infons contain atoms which are all true (given the actual true state) and hence their deviation values lie on the $\vartheta_T(\sigma)$ axis. A similar argument holds for the $\vartheta_F(\sigma)$ axis.

We keep $\iota(\sigma)$ as the {\em informativeness} metric (which is the third dimension in the space, and would be coming out of the page).

We have already seen that on the LHS Floridi used the ratio of the number of false atoms to the length of the state as a metric. That the metrics were different was, as mentioned previously, odd for a straight line scale, but for the current representation it makes sense for those messages that may be true but cannot lie on the horizontal axis because these contain a mixture of true and false atoms. In this case, the number of false atoms can be used as a metric for how far round towards the vertical (false) axis they lie. For the current example, there can be zero, one, or two false atoms, which suggest a suitable set of steps for the angle of each message relative to the horizontal axis. This is depicted in Figure \ref{im}. In the general case, for an infon with $n$ atoms, the angle between each ray would be $\pi/(4n - 2)$ rads.

So far we have only dealt with messages that are true, and we now need to look at how the false infons should be placed in the space. As it happens this is straightforward and follows directly from the relation between the two axes. Being orthogonal they represent contradictory infons: where all the infon lying on the horizontal axis are true and contain only true atoms, those lying on the vertical axis are false and contain only false atoms. In fact they form a ``mirror image'' reflected through a $45^\circ$ line in the space (the dotted line in figure \ref{im}). In this case the distance from the origin is given by the number of worlds in which the infon is false (relative to the true state-of-affairs). Hence the same formula can be used to calculate the distance from the origin. Now we have that the two halves of the space are symmetric around the $45^\circ$ line with the contradictory infons being located at the same relative positions, and therefore having the same distance from the origin (true state).

We are now in a position to put forward a suitable metric to replace that proposed by Floridi.    
The equation of a UC is:

$$x^2 + y^2 = r^2 = 1$$

\noindent extending this to three dimensions (to give a unit sphere) we have: 

$$x^2 + y^2 + z^2 = r^2 = 1$$

\noindent which for the information metric space becomes:

$$\vartheta_T^2(\sigma) +  \vartheta_F^2(\sigma) + \iota^2(\sigma) = 1$$

from which we get the formula for the informativeness metric:
\begin{equation}
\iota(\sigma) = \sqrt{1 - (\vartheta_T^2(\sigma) + \vartheta_F^2(\sigma))}
\label{first}
\end{equation}

At this point we have a clear means whereby we can calculate the deviation of an infon from the true state affairs, and a metric to identify the informativeness of any infon. However, things do not end there. Floridi utilised his metric to quantify the degree of inaccuracy and  degree of vacuity of infons. His approach is very straightforward for inaccuracy, but it is quite complex for the amount of vacuity (requiring the integral of the distance).  It is not difficult to see from Equation \ref{first} that it would be even more so for this space, requiring the double integration of what could be a complex shape. Fortunately we can simplify things significantly by following the approach taken by von Neumann  in the development of QP.

In QP the basic metric is a wave amplitude, which is a periodic (sinusoidal) displacement:  the maximum absolute amplitude is 1 and the minimum is 0, ranging around a UC.  Then if, following von Neumann,  one takes the square of this amplitude one gets the probability of an electron being within that amplitude; then one has

$$P(x) + P(y) = 1$$

\noindent which is a measure rather than a metric.

\begin{figure}[btph]
\centering
\includegraphics[scale=0.8]{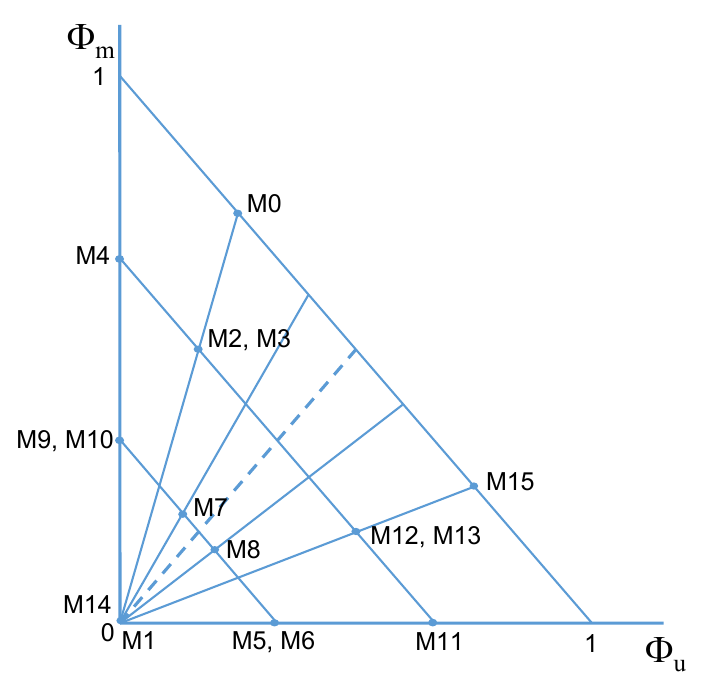}
\caption{\label{is} The distribution of messages in the information measure space.}
\end{figure}

By analogy, 
if we square the metric values $\vartheta_T$, $\vartheta_F$ and $\iota$ we will (as with von Neumann) obtain measure values that generate a new space. The axis obtained from $\vartheta_T$ gives a measure of vacuity or the degree to which one is uninformed (or ignorant).\footnote{Recall that Floridi calculated vacuity by calculating the area under the curve of his relation: squaring a distance has the same `units' as an area.} Let us label that $\Phi_u$. The other axis then is a measure of misinformation, $\Phi_m$ say.\footnote{Or 'Disinformation', but we may ignore that and assume no malicious intent.} And finally, the third axis gives a measure of the informativeness of the infon (and would come out of the page in figure \ref{is}), call that axis $\Phi_i$.
 
Now we have: 

$$\Phi_i + \Phi_m + \Phi_u = 1$$
Or
\begin{equation}
\Phi_i = 1 - (\Phi_m + \Phi_u) = 1 - \Phi_R
\label{mrp}
\end{equation}

\noindent (where $\Phi_R$ is the resultant of $\Phi_m$ and $\Phi_u$). This is MTSI.

As things are set up in the IS for MTSI a contradiction has informativeness of zero (as does a tautology) as required to avoid BCP. However, for reasons presented previously, neither a contradiction nor a tautology lies on either of the the axes of the space since they each contain both true and false atomic infons. Strictly speaking there is no infon on either axis that has a value of zero. On the other hand because (as mentioned previously) a contradiction (or tautology) can contain as many true or false atoms as there are atoms in the messages, each ray not on the axes will contain a version of a contradiction (or tautology) at its zero point, as shown by the $M0$ and $M15$ labels in Figure \ref{is}. As the number of atomic infons increases the points representing a contradiction or tautology will asymptotically approach their respective axes.

The final issue is that according to our measure space all infons and their contradictories have the same value for informativeness. This starts, as we have just seen, with the contradiction and tautology: they are contradictories and both have value zero. As another example consider the atomic infon $x$; it is on the $\Phi_u$ axis and will be true in 50\% of cases. Its contradictory, $x'$, is located on the $\Phi_m$ axis and is false in 50\% of cases. They are images of one another through the $45^\circ$ ($\pi/4$) line and have the same value of informativeness. In one sense it is not surprising that true and false infons have the same value, after all that was also the case for TSSI, but it did not capture this relation for contradictories.
But what is surprising is that this also means that for the true state of affairs, its contradictory will also have a value of 1. That is, for the example used, both $xy$ and $(x' + y')$ yield $\Phi_i = 1$.

The simplest way to show why this is the case is by means of an example. Consider a familiar problem scenario: You are confronted by two doors. Behind one is a million dollars, behind the other is instant death. The doors are protected by two guards, one of whom only tells lies and the other of whom only tell the truth. You have to choose one door, and, obviously, you want to pick the money not the demise.

There are two situations we can explore to make things clear: one in which you know which.guard is which, and one in which you don't. To help you make your decision you are allowed to ask one guard one question. In the first case you can simply ask either guard directly which door you should choose. If you ask Guard 1, then you choose the other door from the one they tell you (because they will misinform you), and if you ask Guard 2, then you choose the door they tell you to take. Simple. Here we see that regardless of which guard you ask, you become equally well informed as to what action to take, because there is a reliable process (logic) that leads you to the correct answer.

In the second situation, which is the one usually used in the puzzle, you can still ask a single question and regardless of which guard you ask, become equally well informed as to which door to choose. In this case you ask either guard what answer the other guard would give. If you ask Guard 1 they will falsely tell you (i.e. misinform you) that Guard 2 would say Door 2. and if you ask Guard 2 they will truthfully tell you that Guard I would say Door 2. So you are now informed that Door 2 is the door you don't want and you can safely choose Door 1. Again, this is because propositional logic provides a robust and reliable means of leading you to the desired result.

We are now in a position to consider how this new approach to semantic information matches up to the desiderata, the BCP and the specified conditions put forward by Floridi (which I believe are correct).\footnote{Floridi states them as conditions for an adequate {\em metric}, but here it is more appropriate to consider them with respect to the proposed information {\em measure}.} 
In section \ref{probs} we saw that TSSI did not meet all the stated conditions for various reasons. Lets look at each one in turn and see how MTSI compares.

\begin{itemize}
\item Clearly Floridi's M.1- M.3 are satisfied in a straightforward manner by the new version of the distance metric, and overcome the problem that M.3 poses for TSSI. 

\item The spirit of M.4 and M.5 is also met straightforwardly, but is adapted to conform to the 2D situation rather than the single dimension of TSSI. For M.4 and M.5, again MTSI achieves what TSSI aimed for but did not provide (i.e. $0 < \Phi_R < 1$). So MTSI gives a robust measure of inaccuracy without the problems highlighted for TSSI. 

\item Then also E.1, E.3 and E.4 have their direct counterpart in MTSI. The fact that MTSI is measure based means that the calculation of quantity of information is simplified with respect to the integral needed for E.2. 

\item Floridi states that ``all atomic messages ought to be assigned the same potential degree of informativeness and therefore, although E.6 indicates that the graph of the model has variable gradient, the rate at which $\iota(\sigma)$ changes with respect to change in$\vartheta(\sigma)$ should be assumed to be uniform, continuous and linear.'' Here again, since the MTSI measure is affine, the aims of E.6 are met exactly, whereas they are only approximated with the quadratic.

\item The information space of MTSI includes all possible messages explicitly, and not just the conjunctive states/ worlds as in TSSI. 

\item Floridi points out that the space he proposed would allow for continuous model to be analysed as well. The same applies here. For the boolean case, as one increases the length of the infon, the space becomes denser but the relative positions of the messages in the space remains the same (e.g. atomic infons will still be true in half the cases etc).

\end{itemize}

There have been a number of challenges to VT and Floridi's TSSI recently. It is beyond the scope of this paper to address these in detail, but a couple of comments are in order. The situation is not helped by the fact that the criticisms deal with different versions, or statements, of VT which makes it harder to address the criticisms. Nonetheless, the paper by Fresco \autocite{fresco17} seems to be based on having a reliable means of shifts from a false to true status. It should be obvious from the foregoing description, and particularly the example of the two doors, that this is not a problem for MTSI. On the other hand, Ferguson \autocite{ferguson15} attacks VT by means of carefully constructed paradoxes whereby VT gives rise to statements that both ``convey information'' and do not ``convey information.'''\footnote{It is not clear why Ferguson utilises the term ``convey information'' rather than simply ``information''. It seems to lead to ambiguity since he says it is synonymous with ``semantic information'' but also refers to it as occurring ``to some degree''.} However one does not need to go to these lengths because Floridi has already provided examples of statements that are information and not information, depending on the {\em accessibility} of the relevant propositions.\autocite[p 50]{floridi10} The example Floridi uses is of a man phoning a garage to get a repair done, stating: ``My wife left the car lights on overnight, and now the battery is flat.'' This statement is false (and hence not information) since it was the man who left the light on. But the relevant part for the mechanic is only that the battery is flat, which is true. One might also suggest in passing that if one were to follow Ferguson here, and reject VT, then unless the term ``misinformation'' is evacuated of all meaning, any statement would both ``convey information'' and ``convey-misinformation''. which does not seem to be an improvement. No doubt the debate on VT will continue, but there is nothing in the current criticism that demands jettisoning it quite yet, and a lot to suggest its continued value.

Finally, and as the observant reader will note, MTSI has the form of an inverse relation. This indicates that it is not the IRP in and of itself that leads to, or entails, the BCP, but the fact that the original version only utilised a single variable. That is, it was at the wrong Level of Abstraction.\autocite{floridi08} TWSI is the same formula as Popper suggested for ``verisimilitude'' \autocite[p 233]{popper63}, and Floridi proposed that ``likeness'' should be used to describe the proximity of messages to the state of interest. This is precisely what MTSI, at least potentially, provides, and complements Cevolani's \autocite{cevolani11} and D'Alfonso's \autocite{dalfonso11} contention that ``truthlikeness'' rather than ``informativeness'' should be the focus of attention. 

To reiterate what was said at the beginning: the sole aim of this paper is to critique TSSI on its own terms and provide an improved version that solves the BCP. The result was MTSI. But that solution gives rise to a number of questions and directions for future research. First, following from D'Alfonso, and Cevolani exploring the possible use of MTSI for verisimilitude is one direction. Also we note that while this theory solves the BCP, it does not address where or how it relates to Floridi's wider Philosophy of Information. This is something that will need to be done, particularly to explore the distribution of messages in the information space when the datum of interest is more abstract, or representative of a different Level of Abstraction \autocite[ch. 3]{floridi11}, or System level analysis \autocite{floridi19} and Informational structural realism \autocite[ch. 15]{floridi11}. As noted above, the more abstract version (Eq.\ref{mrp}) has a similar form to TWSI. This suggests that MTSI may be applicable to a broader spectrum of informational approaches. In particular the GTI framework developed by Burgin \autocite{burgin10} which provides a framework incorporating MTC, probabilistic, and semantic methods, can be explored to assess if and how MTSI fits with that framework.  

\section*{Acknowledgements}
With thanks to Patrick Allo, David Lusseau, and Ulrich Stegmann who gave helpful comments on an earlier version of this paper.

\printbibliography

\end{document}